\documentclass{jfm}

\usepackage{array,mathtools}
\usepackage{natbib}
\usepackage{epsfig,graphics,amssymb,amsmath,latexsym,color}

\def\d{{\rm d}}
\def\x{{\bf x}}
\def\u{{\bf u}}\def\U{{\bf U}}
\def\n{{\bf n}}\def\p{{\partial}}

\title{The other optimal Stokes drag profile}
\author[Thomas D. Montenegro-Johnson and Eric Lauga]{Thomas D. Montenegro-Johnson and Eric Lauga
\thanks{Email address for correspondence: e.lauga@damtp.cam.ac.uk}}
\affiliation{Department of Applied Mathematics and Theoretical Physics, Centre
for Mathematical Sciences, University of Cambridge, Wilberforce Road, Cambridge
CB3 0WA, UK}
\date{\today}

\begin{document}


\maketitle
\begin{abstract}
The lowest drag shape of fixed volume in Stokes flow has been known for some
$40$ years. It is front-back symmetric and similar to an American football with
ends tangent to a cone of $60^\circ$. The analogous convex axisymmetric shape of
fixed surface area, which may be of interest for particle design in chemistry
and colloidal science, is characterized in this paper. This ``other'' optimal
shape has a surface vorticity proportional to the mean surface curvature, which
is used with a local analysis of the flow near the  tip to show that the front
and rear ends are tangent to a cone of angle $30.8^\circ$.  Using the boundary
element method, we numerically represent the shape by expanding its tangent
angle in terms decaying odd Legendre modes, and show that it has $11.3\%$ lower
drag than a sphere of equal surface area, significantly more pronounced than for
the fixed-volume optimal.  
\end{abstract}


\section{Introduction} 
\label{sec:introduction}

Design optimisation is observed throughout nature, and it has long been used in
the private sector for better and more efficient design. Many industries,
particularly aeronautics \citep{jameson1998optimum, alonso2009aircraft} and ship
design \citep{campana2009new}, use fluid dynamical considerations when
optimising shapes \citep{mohammadi2004shape}, such as minimising drag or
maximising lift. These optimisations are usually subject to constraints, for
instance a ship might need to have a fixed volume, or certain shapes might be
too costly to engineer. In nature, too, optimisation is subject to physical
constraints; for example, the beat patterns of spermatozoa
\citep{pironneau1974optimal,lauga2013shape} and cilia
\citep{osterman2011finding,eloylauga12} are subject to energetic constraints on
the molecular motors responsible for the motion. Similarly the shapes of
squirming microorganisms correspond closely to those minimising power
dissipation subject to a maximum curvature \citep{vilfan2012optimal}. Inspired
by nature, shape optimisation can be used to improve the design of biomimetic
artificial swimmers \citep{keaveny2013optimization}.
 
In a classical paper, \citet{pironneau1973optimum} used a variational approach
to show that, in the absence of inertia, the lowest drag shape of fixed volume
is front-back symmetric with constant vorticity on its
surface. Furthermore, the front and rear of the shape taper into cones of
semi-cone angle $60^{\circ}$. This shape was then computed numerically by
\citet{bourot1974numerical} by minimising the drag among a finite family of
shapes. Pironneau's optimal profile has subsequently been observed in the
explosively-launched spores of ascomycete fungi \citep{roper2008explosively}.
The constraint of fixed volume is appropriate in this biological case, as a
certain amount of material is required in order to generate a new organism.

Recently, mechanical optimisations have begun to be examined for more efficient
drug delivery, using material rather than chemical properties to achieve
specific effects \citep{champion2007particle, mitragotri2009physical,
petros2010strategies}. Varying the shape, size, mechanical and surface
properties properties of drugs can significantly impact drug-delivery processes
such as phagocytosis \citep{champion2006role}, whereby macrophages engulf solid
particles. \citet{gratton2008effect} showed that microscopic, slender rod-like
particles with aspect ratio three, corresponding to eccentricity $e = 0.9428$,
were subsumed four times faster than spherical particles of equal volume.

A potentially important constraint for designer drug delivery, and also
biomimetic microrobots, is fixed particle surface area. Indeed, surface area
affects diffusion rates, is associated with toxicity \citep{monteiller2007pro},
and is important in setting the rate of production of reactants for catalysis
and chemical reaction. In this study, we determine the analogous
fixed-surface-area shape to Pironneau's fixed-volume optimal; that is, the
convex, axisymmetric shape that minimises drag subject to fixed surface area.
These constraints restrict us to a family of simply manufactured shapes
resembling Pironneau's optimum, and specifically rule out fractal-like shapes
of finite surface area folded into a very small space. We derive the
fixed-surface-area shape numerically using a representation whereby its tangent angle is
described by a very small number of odd Legendre modes, together with a local
analysis of the equations of fluid motion. We show that this ``other'' optimal
shape is almost twice as slender as the fixed-volume shape,
has a surface vorticity square proportional to the mean surface curvature
\citep{pironneau1973optimum,bourot1974numerical} and near either end is locally
conical about an angle of $30.8^\circ$. 


\section{Mathematical analysis}

Two aspects of the optimisation problem are amenable to an analytical approach,
namely the optimality condition and a local shape analysis near either end of
the body.

\subsection{Optimality condition} 
\label{sec:optimality_cond}

We begin by deriving the shape's optimality condition, following
\citet{pironneau1973optimum}. In Newtonian flows at microscopic scales, the
fluid flow velocity $\mathbf{u}$ driven by a body force per unit volume
$\mathbf{F}$ is governed by the Stokes flow equations,
\begin{equation}
 \boldsymbol{\nabla}p = \mu\nabla^2\mathbf{u} + \mathbf{F} , \quad
\boldsymbol{\nabla}\cdot\mathbf{u} = 0,
\label{eqn:stokes_body_force}
\end{equation}
with pressure $p$ and  viscosity $\mu$. Consider a shape $S$ described by
 $\x$ translating in this fluid at a fixed speed $\U$ in the absence
fluid body forces ($\mathbf{F}=\bf 0$). In the  reference frame of the
body  the velocity satisfies the no-slip boundary condition as
$\u(\x)=\bf 0$ and $\u=-\U$ at infinity. We want to find the
 shape minimizing
the drag. The rate of working of the drag is equal to the rate of viscous
dissipation in the fluid, and so we wish to minimize 
\begin{equation}
D=2\mu \int_V e_{ij} e_{ij} \d V,
\end{equation}
where $V$ refers to the volume outside of the shape and $e_{ij}=(\p u_i / \p x_j
+ \p u_j / \p x_i)/2$.

Perturbing  $S$ into a new shape described by $\x + \alpha \n$, where $\n$ is
the normal to the surface and $\alpha$ is assumed to be a small length
everywhere, we denote the resulting change in the velocity field around the
shape by $\delta \u$. The change in the dissipation is given by 
\begin{equation}
\delta D=\mu  \int_V 
\left(\frac{\p u_i}{ \p x_j} +\frac{ \p u_j }{ \p x_i}\right)
\left(\frac{\p \delta u_i}{ \p x_j} +\frac{ \p \delta u_j }{ \p x_i}\right)
 \d V.
\end{equation}
Using integration by parts and symmetries in the indices we obtain
\begin{equation}\label{3}
\delta D=-2\mu  \int_V 
\delta u_i \left(\frac{\p^2 u_i}{ \p x_j \p x_j} +\frac{ \p^2 u_j }{ \p x_i\p x_j}\right)
 \d V
+2\mu \int_S \delta u_j \left(\frac{\p u_i}{ \p x_j} +\frac{ \p u_j }{ \p x_i}\right) n_i \d S.
\end{equation}
Since $\nabla\cdot \u =0$, we have identically ${ \p_j\p_i u_j }=0$.
The first term in the first integral in equation \eqref{3} is the Laplacian of the
velocity field, which we know from equation \eqref{eqn:stokes_body_force} in the
absence of body forces is given by 
$\mu {\p_j\p_j u_i}  = {\p_i p}.
$ 
Furthermore, a Taylor expansion of the no-slip boundary condition $(\u + \delta
\u) (\x + \alpha \n)=\bf 0 $ allows us to obtain
\begin{equation}\label{5}
\delta \u (\x) = - \alpha \frac{\p \u}{\p n}\cdot
\end{equation}
With these two results, equation \eqref{3} simplifies to 
\begin{equation}\label{6}
\delta D=-2  \int_V 
\delta u_i \frac{\p p}{\p x_i} 
 \d V
-2\mu \int_S  \alpha \frac{\p u_j}{\p n} \left(\frac{\p u_i}{ \p x_j} +\frac{ \p u_j }{ \p x_i}\right) n_i \d S.
\end{equation}
Using integration by parts, the first integral in equation \eqref{6} becomes 
\begin{equation}\label{7}
 \int_V 
-p\frac{\p  \delta u_i}{\p x_i}
 \d V + 
 \int_S p \delta u_i  n_i \d S.
\end{equation}
The first term in equation \eqref{7} is zero because the perturbation flow is
divergence free, while the second term is zero because of equation \eqref{5} and
the divergence-free condition. Because of the no-slip boundary condition for
$\u$ and the fact that it is divergence free, the last integral in equation
\eqref{6} can be further simplified to 
\begin{equation}
\delta D=
-2\mu \int_S  \alpha \frac{\p u_j}{\p n} \frac{ \p u_j }{ \p n}  \d S=
-2\mu \int_S  \alpha \bigg|\frac{\p \u}{\p n} \bigg|^2  \d S.
\end{equation}
In the fixed-surface-area case, we consider a cost function of the form
\begin{equation}
J=D + \lambda \left(\int_{S} \d S-S_0\right),
\end{equation}
where $\lambda$ is the Lagrange multiplier enforcing the constraint of fixed
surface area, $S=S_0$. Variations in $S$ arising
from the perturbation to the shape are given by $2\int\kappa \alpha \d S$
where $\kappa$ is the mean surface curvature, and thus the perturbation to $J$ is given by 
\begin{equation}
\delta J = \delta D + \lambda \delta S 
= -2\mu \int_S  \alpha \bigg|\frac{\p \u}{\p n} \bigg|^2  \d S+ 2\lambda
\int_{S}\kappa {\alpha} \d S
= 2 \int_S  \alpha\left(-\mu  \bigg|\frac{\p \u}{\p n} \bigg|^2  + \lambda \kappa  \right) \d S.
\end{equation}
By setting $\delta J = 0$, the optimality condition is then
\begin{equation}\label{condS}
|\partial \mathbf{u}/\partial {{n}} |^2 \propto \kappa,
\end{equation}
where the first term can also be seen as the norm squared of the surface
vorticity. Note that in the fixed-volume case, the cost function is $J=D +
\lambda (V - V_0)$, and since  $\delta V = \int \alpha \d S $, the optimal
shape has constant surface vorticity \citep{pironneau1973optimum}.
 
\subsection{Local analysis}
\label{sec:local}
\begin{figure}	
\begin{center}
\includegraphics[scale = 0.9]{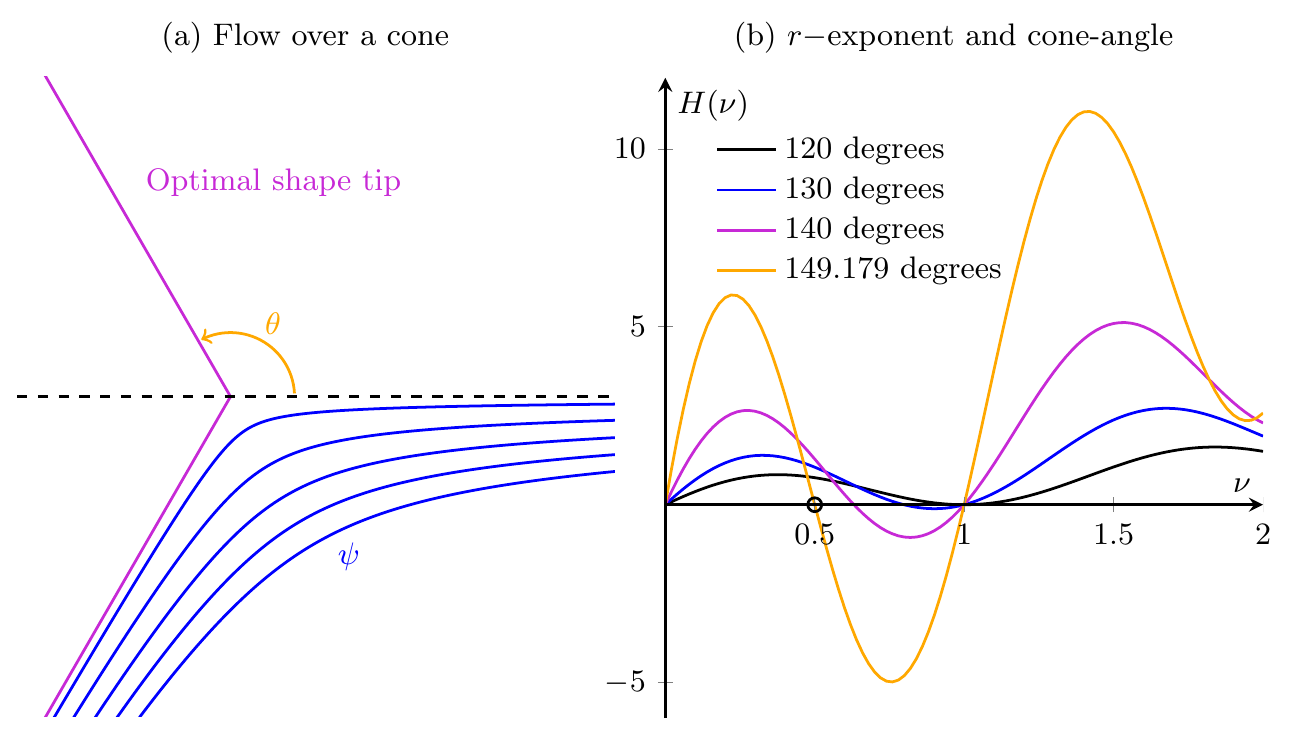}
\end{center}
\caption{(a) Flow over the vertex of a cone illustrating the streamfunction
 for the minimal drag fixed volume case of $\theta_0=120^{\circ}$; (b) The
function $H(\nu)$ in equation \eqref{eq:wakiya_legendre} for a selection of cone
angles, highlighting the minimal drag fixed surface area case of
$\theta_0=149.179^{\circ}$.}
\label{fig:cone_angle}
\end{figure}

As shown by \citet{pironneau1973optimum} and \citet{bourot1974numerical},
optimal shapes have pointy, American football-like ends. The second aspect of
the optimisation problem which can be treated analytically is a derivation of
the optimal angle at either end of the locally-conical shape.

We consider a conical shape and describe the axisymmetric flow around it using
spherical coordinates $(r,\theta)$ with $r$ measured from the tip and $\theta$
from the axis of symmetry in the fluid (figure \ref{fig:cone_angle}a). The cone
is assumed to be located at $\theta=\theta_0$, where $\theta_0$
may be $90^{\circ}$ in the case of a blunt body, and the goal of the analysis is
to derive the value of $\theta_0$ consistent with the optimality condition.
Locally around the cone, the streamfunction $\psi(r,\theta)$ is expected to take
the asymptotic form $\psi = r^n f(\theta)$. The equations and boundary
conditions satisfied by the streamfunction are
\begin{gather}
E^2(E^2\psi) = 0, \,\, E^2 = \frac{\partial^2}{\partial r^2} +
\frac{\sin\theta}{r^2}\frac{\partial}{\partial\theta}\left[\frac{1}{\sin\theta}
\frac{\partial}{\partial\theta}\right], \,\,
\left.\frac{\partial\psi}{\partial\theta}
\right|_{\theta_0} = 0, \,\, \psi(\theta_0) = \psi(0) = 0.
\label{eq:cone_streamfunction}
\end{gather}

In the case of fixed volume, the optimality condition of constant surface
vorticity yields
\begin{equation}
|\partial \mathbf{u}/\partial  n | = |E^2\psi /r \sin\theta| =
\mbox{const.} \quad \Rightarrow \quad n = 3.
\end{equation}
The solution for $f(\theta)$ given in \citet{pironneau1973optimum} does not,
however, satisfy the condition that $\psi(\theta_0) = 0$ (equation
\ref{eq:cone_streamfunction}),  and we re-derive it here.
Plugging the ansatz $\psi = r^3 f(\theta)$ into equation
\eqref{eq:cone_streamfunction} leads to the ordinary differential equation
\begin{equation}
f^{\prime\prime} - \cot\theta f^\prime + 6f = A\cos\theta + B.
\end{equation}
Together with the particular solution, the full solution is then $
\psi = cr^3(at + bt^3 + 1) $ for $t=\cos\theta$. In order to enforce the boundary conditions, the
constants must be $a = -3/2t_0$, $ b = 1/2t_0^3$. Finally, since the streamline
along $\theta = 0$ (or $t = 1$) is split by the cone, we require $f(0)=0$ i.e.
\begin{equation}
\frac{1}{2t_0^3} - \frac{3}{2t_0} + 1 = 0 \quad \Rightarrow \quad (t_0 +
1/2)(t_0 - 1)^2 = 0, 
\end{equation}
and therefore $t_0 = -1/2$ or $\theta_0 = 120^\circ$. The optimal fixed-volume
body thus has locally conical ends with a semi-angle $\theta_{sc}=60^\circ$,
agreeing with the result of \citet{pironneau1973optimum}. The streamlines
associated with this flow are displayed in figure \ref{fig:cone_angle}a.

For the fixed-surface-area shape, the optimality condition depends on the mean
curvature $\kappa$ of the surface (equation \ref{condS}). Note that this means
we expect the vorticity on the surface of our optimal shape to diverge towards
its front and rear ends. At the conical tip, $\kappa \propto 1/r$, and following
the same local analysis, this condition may be written
\begin{equation}
r|\partial \mathbf{u}/\partial n |^2 = r|E^2\psi /r
\sin\theta|^2 = \mbox{const.} \quad \Rightarrow \quad n = 5/2.
\end{equation}
In order to derive the cone angle corresponding this value $n=5/2$, we turn to
\citet{wakiya1976axisymmetric} who considered Stokes flows near the tip of rigid
bodies. In particular, Wakiya showed that, for a given cone angle $\theta_0$,
the  exponent $n$ of the streamfunction satisfying
\eqref{eq:cone_streamfunction} is given by the first nontrivial root (ie, $\nu
\neq 0,1$) of
\begin{equation}
H(\nu) = P_\nu(t_0)\left[t_0 P^{\prime\prime}_\nu(t_0) +
P^{\prime}_\nu(t_0)\right] - t_0\left[P^{\prime}_\nu(t_0)\right]^2,
\label{eq:wakiya_legendre}
\end{equation}
where $\nu + 2 = n$ and $P_\nu$ is the Legendre function of the first kind of,
possibly fractional,
degree $\nu$. The function $H(\nu)$ is plotted for cone angles in figure
\ref{fig:cone_angle}b.  The fixed-volume shape is the 
special case where $\nu = 1$ is a double root of equation
\eqref{eq:wakiya_legendre}. In the case of a fixed-surface constraint, we have
$\nu = 1/2$, and we can  numerically invert equation \eqref{eq:wakiya_legendre}
to obtain the corresponding value of $t_0$. This is illustrated in figure
\ref{fig:cone_angle}b and we obtain  $\theta_0 = 149.179^\circ$ in our
coordinate system, corresponding to a semi-cone angle of 
\begin{equation}\label{sc}
\theta_{sc} \approx 30.8^\circ.
\end{equation}
We thus expect to obtain a more slender optimal shape than in the fixed-volume
case.

\section{Numerical shape optimisation}
\label{sec:modeling}

We now proceed with a computational approach in order to derive the optimal
shape and confirm the validity of our local analysis. The fundamental solution
to the Stokes flow equations \eqref{eqn:stokes_body_force} driven point a point
force $\mathbf{f}\delta(\mathbf{x} - \mathbf{y})$ located at $\mathbf{y}$ is
given by the stokeslet tensor $S_{ij}(\mathbf{x}, \mathbf{y})$,
\begin{equation}
u_i(\mathbf{x}) = S_{ij}(\mathbf{x},\mathbf{y})f_j, \quad
S_{ij}(\mathbf{x},\mathbf{y}) =  \frac{1}{8\pi}\left(\frac{\delta_{ij}}{r} +
\frac{r_i r_j}{r^3} \right),
\end{equation}
where $r_i = x_i - y_i$ and $r^2 = r_1^2 + r_2^2 + r_3^2$. For axisymmetric
flow, the Green's function corresponding to a ring of point forces
$M_{\alpha,\beta}(\mathbf{x} - \mathbf{x}_0)$ can then be obtained by
integrating the Stokeslet in the azimuthal direction
\citep{pozrikidis1992boundary},
\begin{gather}
M_{xx} = 2k\left(\frac{\sigma_0}{\sigma}
\right)^{1/2}\left(F + \frac{\hat{x}^2}{\hat{r}^2}E \right),\quad  
M_{x\sigma} = -k\frac{\hat{x}}{\sigma_0\sigma^{1/2}}
\left[F - (\sigma^2 - \sigma_0^2 +\hat{x}^2)\frac{E}{\hat{r}^2} \right],
\nonumber \\
M_{\sigma x} = k\frac{\hat{x}}{\sigma}
\left(\frac{\sigma_0}{\sigma}\right)^{1/2} \left[F + (\sigma^2 - \sigma_0^2 -
\hat{x}^2)\frac{E}{\hat{r}^2} \right], \\
M_{\sigma\sigma} =
\frac{k}{\sigma_0\sigma}\left(\frac{\sigma_0}{\sigma}\right)^{1/2}\left\{(\sigma_0^2
+ \sigma^2 + 2\hat{x}^2)F - \left[2\hat{x}^4 + 3\hat{x}^2(\sigma_0^2 + \sigma^2) +
(\sigma^2 - \sigma_0^2)^2\right]\frac{E}{\hat{r}^2}\right\}. \nonumber
\label{eqn:axisymmetric_stokeslet}
\end{gather}
Here, $F$ and $E$ are complete elliptic integrals of the first kind of argument
$k$, 
\begin{gather}
F(k) = \int_0^{\pi/2}\frac{\mathrm{d}\omega}{(1 - k^2 \cos^2\omega)^{1/2}}, \quad
E(k) = \int_0^{\pi/2}(1 - k^2 \cos^2\omega)^{1/2}\,\mathrm{d}\omega ,  
\quad  
\end{gather}
where  $k^2 = {4\sigma\sigma_0}/[{\hat{x}^2 + (\sigma + \sigma_0)^2}]$,  while $\hat{x} = x - x_0$ and $r^2 = \hat{x}^2 + (\sigma - \sigma_0)^2$ are 
axial and radial coordinates $(x,\sigma)$.

The Newtonian fluid flow around our axisymmetric body can then be reduced to a
boundary integral of ring forces taken over the meridional plane,
\begin{equation}
u_\alpha(\mathbf{y}) = \frac{1}{8\pi}\int_C
M_{\alpha\beta}(\mathbf{x},\mathbf{y})f_\beta(\mathbf{y})\,\mathrm{d}l,
\label{eqn:BIE}
\end{equation}
where the double layer potential has been eliminated (rigid-body motion). The
meridional plane of the rigid body is discretised into straight-line segments of
constant force per unit length; i.e. the components $f_x,f_\sigma$ in equation
\eqref{eqn:BIE} are constant over each element $s_i$. The velocity of the body
is specified at the midpoint of each element. Numerical evaluation of each
non-singular line integral is performed with four-point Gaussian quadrature,
while singular integrals have the stokeslet singularity removed and integrated
analytically. The axisymmetric Green's function is evaluated using BEMLIB
\citep{pozrikidis2002practical}. The drag on the body is then calculated by
summing the products of the area of each element by the calculated traction.

By computing the drag, we can create a series of profiles with decreasing drag
relative to the equivalent sphere. For this, we adopt the approach of
\citet{bourot1974numerical} whereby the drag is minimised within a family of
parametrized shapes, a method  also used by \citet{zabarankin2013minimum}  to
determine optimal profiles in elastic media. By symmetry of the underlying
equations, we restrict our attention to axisymmetric shapes with front-back
symmetry \citep{pironneau1973optimum}.  An efficient description of such shapes
may be obtained by expanding their tangent angle, $\phi(s)$, in odd Legendre
polynomials, $P_{2n-1}$,
\begin{equation}\label{Legendre}
\phi(s) = \sum_{n = 1}^{N} A_n P_{2n-1}(s),
\end{equation}
for $s$ arclength along the meridional plane. The $(x,\sigma)$ positions of the
shape are then
\begin{equation}
x = \int_C \cos[\phi(s)]\, \mathrm{d}s, \quad \sigma = \int_C \sin[\phi(s)]\,
\mathrm{d}s.
\label{eq:shapes}
\end{equation}
Integration of equation \eqref{eq:shapes} is performed numerically by Simpson's
rule over an ultrafine mesh, prior to the descritisation of the boundary into
straight-line elements. The volume and surface area of the resultant shape are
calculated by summing conical frustra defined by the ultrafine mesh.
Optimisation is performed for $n = 1\to N$ sequentially, and the converged
optimal coefficients for the $n=N-1$ shape are used as an initial guess for the
$n=N$ shape, with $A_N$ initially set to zero. Optimization is
carried out using the \textsf{fmincon} function in \textsf{Matlab}, employing a
sequential quadratic programming (SQP) algorithm. In constrained optimizations,
the sum of the Legendre coefficients is forced to equal the analytically-calculated tip angle at every iterative step. Hessians for each step are computed
numerically via finite differences.

In order to validate the numerical procedure, we compare the drag on ellipsoids of
varying eccentricity and fixed volume as calculated by our boundary element code
against the analytical result of \citet{chwang1975hydromechanics} 
\begin{equation}
\mbox{Drag} = 6\pi\mu a U C_{f1},\quad C_{f1} = \frac{8}{3}e^3\left[-2e + (1 +
e^2)\log \frac{1+e}{1-e} \right]^{-1}.
\label{eq:chwang_drag}
\end{equation}
The percentage relative errors in the drag between our work and
\citet{chwang1975hydromechanics}  for 100, 200, 400 and 800 mesh elements are,
respectively, 0.0102\%, 0.00256\%,  0.000641\%,  and 0.000160\%. 
The drag of successive optima for increasing numbers of modes converges very
quickly. As a very high degree of accuracy is  required to fully resolve the
optimisation for higher numbers of modes,  we discretise the shape into $800$
elements. As a further validation to ensure that forces were sufficiently
resolved around the conical ends of spindle-like bodies, the drag force on a
spindle of $120^{\circ}$ was found to be in agreement with that provided by
\citet{wakiya1976axisymmetric}.

\begin{figure}	
\begin{center}
\includegraphics[scale = 0.9]{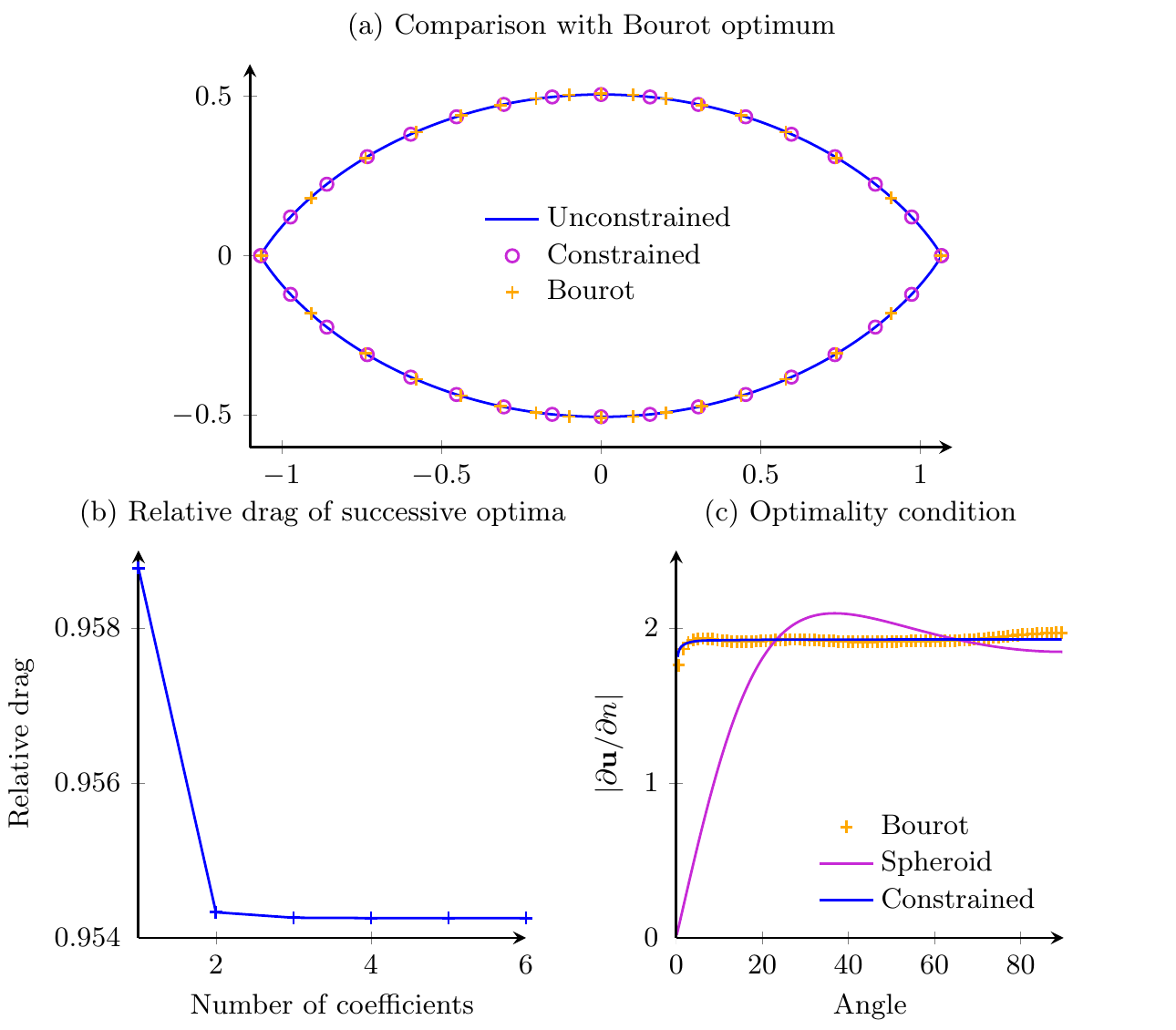}
\end{center}
\caption{(a): The lowest drag shape of fixed volume as calculated by constrained
(fixed front/rear angle) and unconstrained optimisation, together with the shape
originally given by \citet{bourot1974numerical}; (b): The drag relative to a
sphere of unit volume of succesive optima in found by our constrained
optimisation and (c): Pironneau's optimality condition over the surface of
Bourot's optimum, the optimal prolate spheroid of fixed volume and our
constrained optimum.}
\label{fig:vol_opt}
\end{figure}

As a second numerical check, we verify that expanding the tangent angle in terms
of Legendre modes provides an efficient description of the lowest drag shape of
fixed volume. We consider both an optimisation scheme whereby the front and rear
semi-cone angles are constrained to $60^{\circ}$ (see above), and a fully
unconstrained optimisation. Figure \ref{fig:vol_opt}a shows our constrained and
unconstrained optima for $6$ Legendre modes in comparison to the shape given by
\citet{bourot1974numerical}, described by $10$ coefficients of the equation
\begin{equation}
\frac{r(\theta)}{\lambda} = 2 - \frac{2}{\sqrt{3}}\sin\theta + \sum_{n=1}^{10}
B_n \sin^{n+1}\theta.
\end{equation}
The values for the coefficients given by \citet{bourot1974numerical} are
truncated to $4$ s.f., and  their shape  does not quite have the same volume as
a sphere of unit radius. Nonetheless, it is clear that our description of the
tangent angle by odd Legendre modes yields the optimal shape quickly, see
e.g.~the convergence of the relative drag in figure \ref{fig:vol_opt}b. This
fast convergence is why a large number of elements ($N = 800$) are required in
the unconstrained optimisation for the angle at the tip to converge to
$60^{\circ}$. Note that the optimal shape for only $3$ Legendre modes is almost
indiscernible from our final optimal. After $6$ modes in the unconstrained
optimisation, the calculated tip angle converged with $2\%$ relative error. The
relative drag of the optimal shape (ratio of drag to that of sphere of
equivalent volume) is $D_r = 0.9542549$, in agreement with the result of
Bourot, $D_r =0.95425$. Pironneau's optimality condition of a constant surface
vorticity is also confirmed to be valid   in  figure \ref{fig:vol_opt}c. 

\begin{figure}	
\begin{center}
\includegraphics[scale = 0.9]{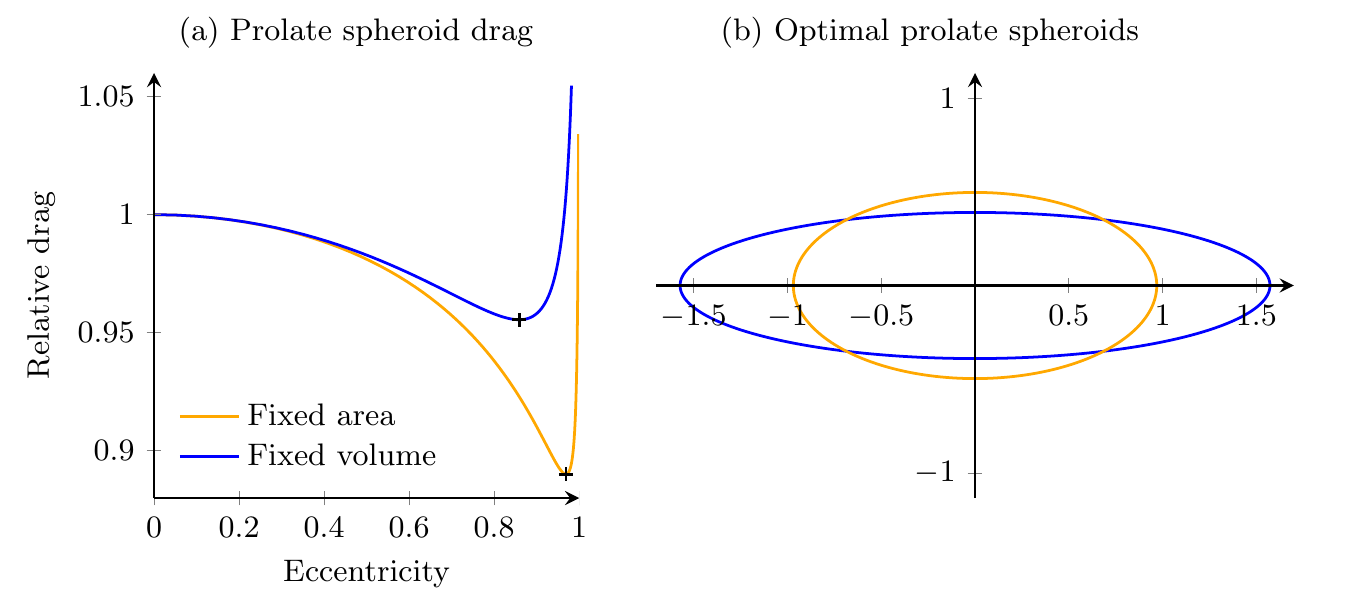}\end{center}
\caption{(a): The drag on prolate spheroids of fixed volume (dark blue) and
surface area (light yellow) relative to their equivalent spheres as a function
of eccentricity. The minimum drag prolate spheroids are marked with plusses;
(b): Shape of the minimum drag prolate spheroids of fixed volume and
surface area, normalised to unit volume.}
\label{fig:ellipsoid_test}
\end{figure}

In order to provide a base-line comparison for the numerical shapes, the drag
was first minimised for among the family of prolate spheroids of varying
eccentricity for both fixed volume and fixed surface area using the analytical
result in equation \eqref{eq:chwang_drag} and our boundary element code. The
computational code  achieves results identical to within $10^{-8}$ of the
analytical ones and the optimal spheroids have eccentricities  $e = 0.8588$
(aspect ratio 1.952) in the fixed-volume case  and $0.9688$ (aspect ratio
4.037) in the fixed-surface-area case. We show these results in figure
\ref{fig:ellipsoid_test}, and they allow us to anticipate on two key facts of
drag minimisation for fixed surface areas. First, we can predict that the
lowest drag shape of fixed surface area is likely to be more slender than its
fixed-volume counterpart (this was already anticipated in the local 
analysis above). Furthermore, we can expect that the reduction in drag of the
fixed-surface-area optimum relative to its equivalent sphere will be
greater  than the fixed volume counterpart; indeed, for spheroids, the reduction
in drag is $ 11\%$ for fixed surface area but only $4.45\%$ for fixed
volume.

\begin{figure}	
\begin{center}
\includegraphics[scale = 0.9]{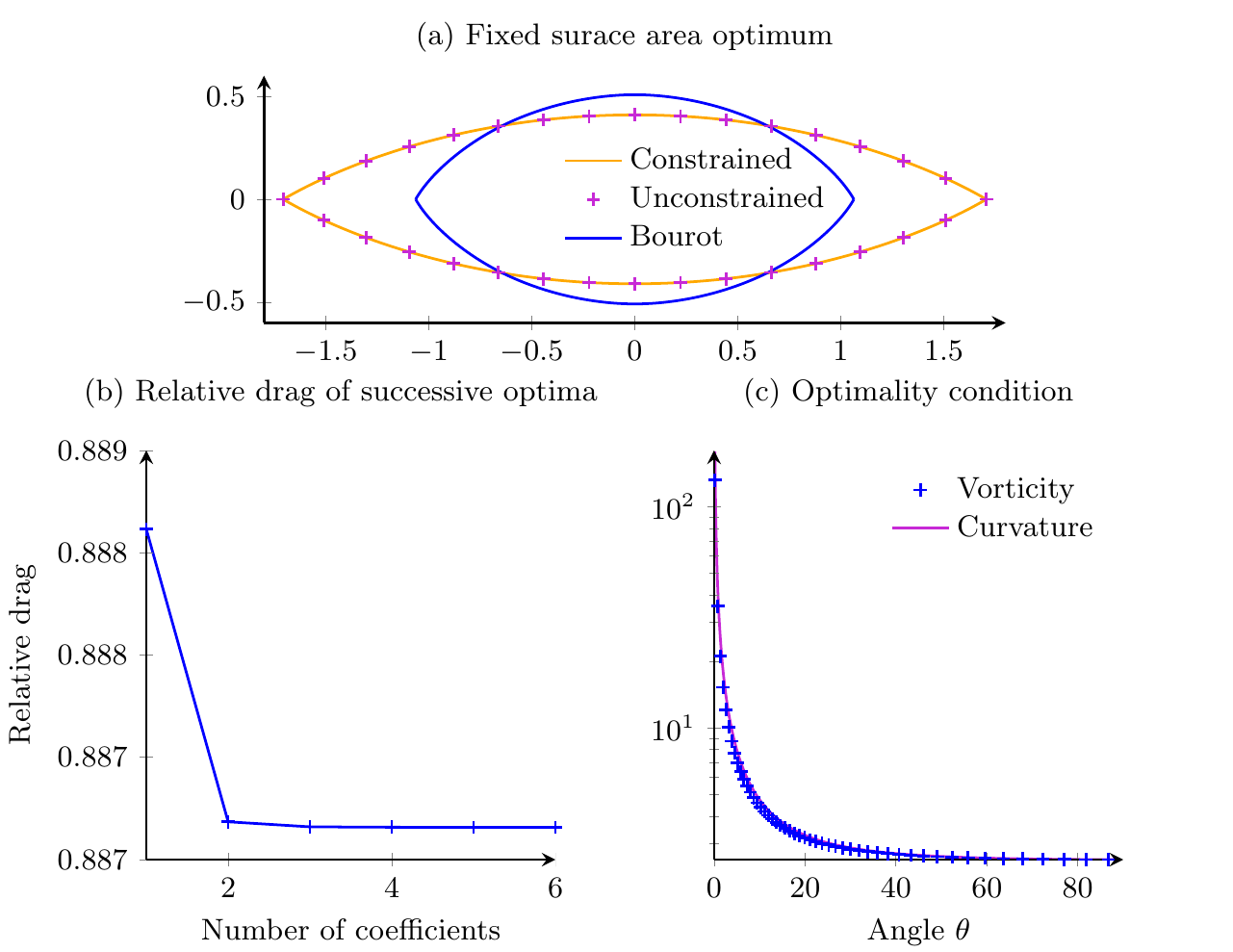}
\end{center}
\caption{(a): The lowest drag shape of fixed surface as calculated by
unconstrained ($+$) and constrained 
optimisation (i.e.~fixed tip angles, light yellow line), in comparison with the fixed volume case (dark blue line). Both shapes are normalized to have unit volume; (b): The drag
relative to a sphere of unit surface area for successive optima using 
constrained optimisation; (c): Vorticity ($+$) along the surface of the
optimal shape of fixed surface area shown with its mean curvature, scaled by
 the ratio of vorticity to curvature at $\theta = 90^\circ$ (solid purple
line).}
\label{fig:area_opt}
\end{figure}

We now compute the fixed-surface-area analogue of Pironneau's
shape using both unconstrained and constrained (i.e.~fixed tip angle)
optimisation. After $6$ modes in the unconstrained optimisation, the calculated
tip angle converged to the analytically calculated optimal of $30.8^\circ$
(equation \ref{sc}) with a $1\%$ relative error. The optimal shape found with
both optimisation approaches is displayed in figure \ref{fig:area_opt}a. The
unconstrained and constrained results are indistinguishable. We
also included in figure \ref{fig:area_opt}a the  fixed-volume optimal, with all
shape  volumes normalised to $1$ in order to facilitate comparison. As
anticipated, the fixed-area optimal is much more slender than the best shape of
fixed volume (aspect ratio of $4.162$ against $2.109$, which are both slightly
above the aspect ratios of the optimal spheroids). As with the volume
constraint, the drag converges here to its optimal value with very few modes,
which is illustrated in figure \ref{fig:area_opt}b. The first six unconstrained
coefficients describing the tangent angle in equation~\eqref{Legendre} are
\begin{align}
A_1 =& -0.47956, &A_2 =& -0.035208, &A_3 =& -0.0096958,\nonumber \\
A_4 =& -0.0042400, &A_5 =& -0.0022198, &A_6 =& -0.0012228.
\end{align}

How much better than the equivalent sphere is the optimal shape? The drag ratio
we obtain is  equal to $D_r = 0.8871581$, so a drag reduction  of about 11.3\%.
This is significantly greater than the $4.57\%$ reduction for the case of fixed
volume. Finally we are able to test numerically the optimality criterion
derived in equation~\eqref{condS}. We plot in \ref{fig:area_opt}c both the value
of the flow vorticity on the surface and the mean surface curvature, $\kappa$.
As can be seen, our computational results show a perfect proportionally between
both, confirming the validity of the local analytical criterion discussed in
Sec.~\ref{sec:optimality_cond}.


\section{Conclusion}
\label{sec:discussion}

Shape optimisation at small scales is of potential significance for problems in
chemistry and colloidal science. In this work, we used the boundary element
method together with a mathematical analysis to determine computationally the
convex, axisymmetric shape that minimises drag in Stokes flow subject to the
constraint of a fixed surface area.  The shape is about twice as slender as the
fixed volume case, and the front and rear ends of the shape are tangent to a
cone of semi-angle $\approx 30.8^{\circ}$.  This was borne out by minimisation
of the drag with respect to coefficients of an efficient representation of the
shape's tangent angle in terms of Legendre modes. Although we did not
mathematically prove that the shape we obtained is globally optimal, or in fact
unique, among all axisymmetric convex ones, since it quickly converges to it
from all initial conditions, we conjecture that it is.

From a practical standpoint, while the optimality conditions for the fixed
volume and surface area shapes entail the vorticity on the surface of these
shapes, the drag turns out to be somewhat insensitive to details of surface
vorticity. In fact, the optimal prolate spheroids of fixed volume and surface
area both have zero vorticity at the front and rear ends, but only exhibit
$0.1\%$ and $0.3\%$ more drag than their respective full optima. This
insensitivity engenders the notion of ``nearly optimal design'' in Stokes flow;
optimisation of simple shapes can be effective and helpful to experimental
design.
 
\subsection*{Acknowledgements}
\label{subsec:acknowledgements}
 This work was funded in part by a European Union Marie Curie CIG to EL and a
Royal Commission for the Exhibition of 1851 Research Fellowship to TDMJ.

\bibliographystyle{jfm}
\bibliography{refs}

\end{document}